\documentclass[aps,prx,reprint,superscriptaddress,longbibliography]{revtex4-2}

\usepackage{amsmath,amssymb,mathtools,bm}
\usepackage{braket}
\usepackage{graphicx}
\usepackage{hyperref}
\usepackage{xcolor}
\usepackage{enumitem}
\usepackage{microtype}

\hypersetup{colorlinks=true,linkcolor=blue,citecolor=blue,urlcolor=blue}

\newcommand{\ii}{\mathrm{i}}
\newcommand{\dd}{\mathrm{d}}
\newcommand{\Tr}{\operatorname{Tr}}

\newtheorem{theorem}{Theorem}

\newtheorem{corollary}{Corollary}

\begin{document}

\title{Interface Capacity and Architectural Replenishment Determine Entanglement-Generation Speed in Quantum Networks}

\author{Shi-Ju Ran}
\email{sjran@cnu.edu.cn}
\affiliation{Center for Quantum Physics and Intelligent Sciences, Department of Physics, Capital Normal University, Beijing 100048, China}

\date{\today}

\begin{abstract}
We show that entanglement-generation speed across a fixed network interface is governed by two distinct resources: the entangling capacity of the interface itself and the ability of the surrounding architecture to replenish it with fresh degrees of freedom. For fermionic Gaussian dynamics, we derive the coefficient-sharp bound $\sum_k|\dot\theta_k|\leq\frac12\|K_{AB}\|_*$ on the collective speed of the canonical entanglement angles. Explicit Ising-chain rematching trajectories saturate this bound, thereby certifying exact minimum interaction times under the stated control model. Beyond the Gaussian setting, exhaustive optimization of the complete $N=8$ tree--tree family shows that, at fixed interface capacity, first-layer entanglement, connectedness, and edge budget, the saturation depth is exactly classified by rooted architecture. With higher-resolution $x$-only control, variational entanglement-enhancing-field (VEEF) optimization reaches the numerically resolved fast-$X$ optimum in a two-channel benchmark. Across all 21 symmetry-reduced rooted orbits, a pre-specified two-time VEEF growth diagnostic recovers the complete replenishment partition directly from optimized dynamics. Interface capacity therefore sets how much entangling flux is available, whereas architecture determines whether fresh degrees of freedom can continually replenish the interface and sustain repeated use of that capacity.
\end{abstract}

\maketitle

\section{Introduction}
\label{sec:intro}

Entanglement distribution is a basic primitive for quantum communication and distributed quantum computation~\cite{Cirac1997,Briegel1998,Cirac1999}. In fixed-hardware processors and quantum networks, however, the relevant question is not only whether entanglement can be generated, but how quickly it can be generated across a repeatedly used interface. The interactions that cross a bipartition set an immediate entangling resource, while the surrounding architecture must move or rearrange degrees of freedom so that the same interface can continue to create new entanglement. This motivates the central question of this work: what sets entanglement-generation speed when the physical interface is fixed but internal routing remains dynamical and architecture constrained?

The entangling capabilities and capacities of bipartite Hamiltonians under local assistance are well established~\cite{Dur2001,Bennett2003,Childs2003}, and entangling-rate, locality, and flow bounds constrain how correlations and entropy can grow~\cite{LiebRobinson1972,BravyiHastingsVerstraete2006,EisertOsborne2006,Bravyi2007,VanAcoleyen2013,Cubitt2005}. Complementary work on time-optimal and fast-local control characterizes minimum-time spin dynamics and entanglement generation~\cite{Khaneja2001,Romano2007,Caneva2009,Koch2022,Malvetti2024,MalvettiVanDamme2024}, while architecture-sensitive routing bounds show how connectivity and bottlenecks constrain quantum information and entanglement transport~\cite{Eldredge2020,Bapat2023,Devulapalli2026}. These strands leave a controlled question: if the entire cross-cut interaction is held fixed, how much of the attainable speed is set by that interface itself, and how much by the internal architecture's ability to reuse it?

We organize the problem around two resources: \emph{interface capacity}, the entangling flux supplied by the fixed cross-cut interaction, and \emph{architecture-dependent replenishment}, the ability of internal dynamics to keep presenting fresh degrees of freedom to that interface. We fix the interaction graph and bipartition, allow arbitrarily fast on-site rotations, and keep all two-site interactions finite-strength and architecture constrained. In the fermionic Gaussian sector, the capacity side is exact: we prove the coefficient-sharp bound $\sum_k|\dot\theta_k|\leq\|K_{AB}\|_*/2$, whose integral gives a many-mode quantum speed limit. Kicked-Ising entanglement protocols provide the relevant staged growth trajectories~\cite{Mishra2015}. The Majorana-rematching representation used here shows that explicit chain constructions saturate the bound and therefore attain the minimum interaction times allowed by the stated control model.

With the interface held fixed, the remaining speed variation is architectural. Across the complete $N=8$ tree--tree family, the topology relative to the fixed interface exactly determines the efficiency of replenishment and hence the entanglement-generation speed. We then allow higher-resolution $x$-only control. VEEF control~\cite{Lu2021,Lu2024} reaches the numerically resolved fast-$X$ optimum in a calibrated architecture, while a pre-specified two-time VEEF growth diagnostic identifies the complete $L_*=2/3$ rooted replenishment partition across all 21 reduced orbits. The continuous dynamics therefore provide an inverse probe of the same rooted replenishment structure found by exhaustive fixed-layer optimization.

\section{Exact Gaussian Interface Capacity}
\label{sec:gaussian_capacity}

For a bipartite Hamiltonian $H(t)=H_A(t)+H_B(t)+H_\partial(t)$, only the cross-boundary term can directly change the bipartite entanglement. The central question is therefore not merely how many interaction terms cross the cut, but how rapidly the fixed physical interface can move the entanglement coordinates. In the fermionic Gaussian sector this question admits an exact answer.

\subsection{Interface capacity and Gaussian speed limit}
\label{sec:gaussian_bound}

Let subsystem $A$ contain $N_A$ fermionic modes and subsystem $B$ contain $N_B$ modes, and set
\begin{equation}
m=\min(N_A,N_B).
\end{equation}
For a pure fermionic Gaussian state, local Gaussian transformations bring the bipartite state to a product of $m$ canonical mode pairs~\cite{BoteroReznikFermion2004,BravyiFermionic2005},
\begin{equation}
\ket{\Psi}\simeq
\bigotimes_{k=1}^{m}
\left(\cos\theta_k\ket{0_k0_k}+\sin\theta_k\ket{1_k1_k}\right)
\otimes\ket{\mathrm{local}},
\label{eq:canonical_gaussian_pairs}
\end{equation}
with $0\leq\theta_k\leq\pi/4$. Thus $\theta_k=0$ denotes a product mode pair and $\theta_k=\pi/4$ a maximally entangled one.

Write a quadratic Majorana Hamiltonian as~\cite{BravyiFermionic2005}
\begin{equation}
H(t)=\frac{\ii}{4}\,\gamma^T K(t)\gamma,
\qquad
K(t)=-K^T(t),
\label{eq:majorana_hamiltonian}
\end{equation}
and order the Majoranas according to the bipartition,
\begin{equation}
K(t)=
\begin{pmatrix}
K_A(t) & K_{AB}(t)\\
-K_{AB}^T(t) & K_B(t)
\end{pmatrix}.
\label{eq:majorana_blocks}
\end{equation}
The physical cross-boundary block $K_{AB}(t)$ defines the instantaneous Gaussian interface capacity
\begin{equation}
\Lambda_\partial(t)
\equiv \frac12\|K_{AB}(t)\|_*,
\label{eq:interface_capacity}
\end{equation}
where $\|\cdot\|_*$ is the Schatten $1$-norm. If the singular values of $K_{AB}(t)$ are time independent, we write the constant capacity as $\Lambda_\partial$. With the normalization in Eq.~\eqref{eq:majorana_hamiltonian}, a nearest-neighbor Ising bond of strength $J$ contributes one singular value $2|J|$.

\begin{theorem}[Gaussian angle-flow bound]
\label{thm:gaussian_angle_flow}
For a pure fermionic Gaussian state evolving under Eq.~\eqref{eq:majorana_hamiltonian}, the canonical entanglement angles obey, at every regular point and hence almost everywhere in time,
\begin{equation}
\sum_{k=1}^{m}|\dot\theta_k(t)|
\leq \frac12\|K_{AB}(t)\|_*
=\Lambda_\partial(t).
\label{eq:angle_speed_theorem}
\end{equation}
\end{theorem}

The bound follows by decomposing $K_{AB}$ into singular channels and projecting each channel onto the canonical two-dimensional Majorana planes. For a singular channel of weight $\lambda_\alpha$, Cauchy--Schwarz bounds its total contribution to $\sum_k|\dot\theta_k|$ by $\lambda_\alpha$. Summing over channels gives Eq.~\eqref{eq:angle_speed_theorem}. Local quadratic frame changes act by left and right orthogonal rotations of $K_{AB}$ and therefore preserve its singular values. The full derivation is given in Appendix~\ref{app:gaussian_angle_flow_proof}.

Eq.~\eqref{eq:angle_speed_theorem} differs from general entropy-rate and entanglement-speed bounds~\cite{Bravyi2007,VanAcoleyen2013,Pandey2024}, fast-bipartition-local Schmidt-variable bounds~\cite{Malvetti2024,MalvettiVanDamme2024}, and boundary-current bounds for number-conserving free fermions~\cite{Hamazaki2024}. Here the bound resolves the collective motion of all Gaussian canonical angles and retains the physical cross-boundary Majorana generator. To our knowledge, this coefficient-sharp nuclear-norm bound together with the saturation criterion below is not contained in those formulations.

Define the total canonical entanglement coordinate
\begin{equation}
\Theta(t)=\sum_{k=1}^{m}\theta_k(t).
\end{equation}
Theorem~\ref{thm:gaussian_angle_flow} immediately gives the integrated speed limit.

\begin{corollary}[Gaussian quantum speed limit]
\label{cor:gaussian_qsl}
For any Gaussian trajectory beginning at a product state,
\begin{equation}
\Theta(T)
\leq \int_0^T\dd t\,\Lambda_\partial(t).
\label{eq:integrated_capacity}
\end{equation}
If the interface capacity is constant and the target contains $q$ maximally entangled canonical mode pairs, then
\begin{equation}
T\geq \frac{q\pi}{4\Lambda_\partial}.
\label{eq:qsl_q_modes}
\end{equation}
For maximal bipartite Gaussian entanglement, $q=m$.
\end{corollary}

To separate available capacity from realized flux, define the instantaneous interface utilization
\begin{equation}
u(t)=
\frac{\sum_k|\dot\theta_k(t)|}{\Lambda_\partial(t)}
\leq1,
\label{eq:utilization}
\end{equation}
whenever $\Lambda_\partial(t)>0$. Equality in the integrated bound requires two conditions almost everywhere: the target angles do not backtrack, $\dot\theta_k(t)\geq0$, and the interface is fully utilized, $u(t)=1$. The first avoids wasting canonical-angle motion, while the second requires the admissible dynamics to keep the physical boundary channels aligned with fresh entangling directions. This distinction between interface capacity and repeated saturability is the organizing principle for the examples below.

\subsection{Attainability in uniform Ising chains}
\label{sec:optimal_routing}

\begin{figure}[t]
\centering
\includegraphics[width=\columnwidth]{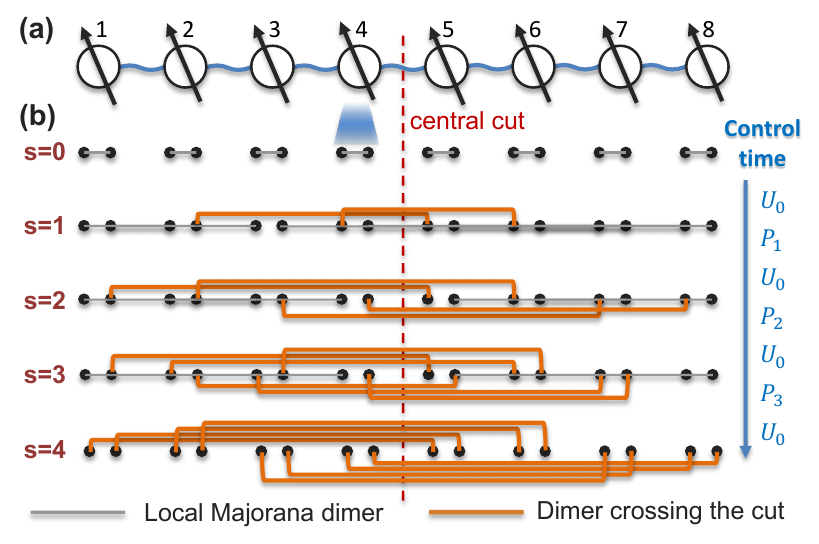}
\caption{\textbf{Exact interface-capacity saturation in an $N=8$ open Ising chain.} The Majorana-rematching trajectory repeatedly presents a fresh canonical pair to the single physical interface. After stage $s$, $2s$ Majorana dimers cross the cut, corresponding to $s$ maximally entangled fermionic mode pairs. The protocol therefore realizes monotonic canonical-angle growth with unit interface utilization and reaches $S_A=4\ln2$ at $T=4t_0$, saturating Eq.~\eqref{eq:obc_lower_bound}.}
\label{fig:obc_majorana_matching}
\end{figure}

Uniform nearest-neighbor Ising chains provide explicit attainability tests of Theorem~\ref{thm:gaussian_angle_flow}. For later use, define the uniform Ising evolution
\begin{equation}
H_G=J\sum_{(i,j)\in E}Z_iZ_j,
\qquad
U_G(t)=e^{-\ii H_Gt}.
\label{eq:uniform_ising_evolution}
\end{equation}
The special interaction time
\begin{equation}
t_0\equiv\frac{\pi}{4|J|}
\label{eq:clifford_time}
\end{equation}
is the \emph{Clifford time}. We call $U_G(t_0)$ one physical Ising layer. Starting from $\ket{+}^{\otimes N}$, we allow arbitrarily fast on-site $X$ rotations,
\begin{equation}
H(t)=H_G+\sum_j h_j^x(t)X_j.
\label{eq:ising_control}
\end{equation}
With the Jordan--Wigner Majoranas~\cite{JordanWigner1928,LiebSchultzMattis1961}
\begin{equation}
a_j=\left(\prod_{\ell<j}X_\ell\right)Z_j,
\qquad
b_j=\left(\prod_{\ell<j}X_\ell\right)Y_j,
\label{eq:jw_majoranas_main}
\end{equation}
one has $X_j=\ii a_jb_j$ and $Z_jZ_{j+1}=\ii b_ja_{j+1}$, so the nearest-neighbor drift and the allowed controls are Gaussian and Theorem~\ref{thm:gaussian_angle_flow} applies directly. For an open chain, one bond crosses the central cut, whereas an even periodic chain has two boundary channels. Hence
\begin{align}
\Lambda_\partial^{\rm OBC}&=|J|,
& T&\geq mt_0,
\label{eq:obc_lower_bound}\\
\Lambda_\partial^{\rm PBC}&=2|J|,
& T&\geq\frac{mt_0}{2},
\label{eq:pbc_lower_bound}
\end{align}
where $m=\lfloor N/2\rfloor$ for the open chain and $N=2m$ for the periodic chain. Known kicked-Ising protocols exhibit the corresponding staged entanglement growth~\cite{Mishra2015}. In the Majorana-rematching representation developed here, the open- and periodic-chain constructions have monotonic canonical-angle growth and unit utilization $u(t)=1$. They therefore attain the interface bounds exactly,
\begin{align}
T_{*,\mathrm{OBC}}&=mt_0
=\frac{m\pi}{4|J|},
\label{eq:TOBC}\\
T_{*,\mathrm{PBC}}&=\frac{mt_0}{2}
=\frac{m\pi}{8|J|}
=\frac12T_{*,\mathrm{OBC}}.
\label{eq:TPBC}
\end{align}
Thus the factor-of-two periodic-chain speedup follows directly from doubled interface capacity together with exact saturation. The explicit rematching construction is given in Appendix~\ref{app:majorana_matching}. Its role here is to certify attainability of the Sec.~\ref{sec:gaussian_capacity} bound. Figure~\ref{fig:obc_majorana_matching} illustrates the saturation mechanism for an $N=8$ open chain.

\section{Rooted Architecture Controls Interface Replenishment}
\label{sec:fixed_interface_ensemble}

The chain calibration shows that the interface bound of Sec.~\ref{sec:gaussian_capacity} can be sustained when the surrounding architecture replenishes the boundary perfectly. We now keep the cross-cut interaction fixed and ask how changing only the internal architecture alters that repeated utilization. For general Ising graphs the Gaussian theorem need not apply because non-nearest-neighbor $ZZ$ couplings generally acquire Jordan--Wigner strings~\cite{JordanWigner1928,LiebSchultzMattis1961}. At the Clifford time $t_0$ defined in Eq.~\eqref{eq:clifford_time}, standard graph-state cut-rank and stabilizer-entanglement results~\cite{Hein2004,Fattal2004}, together with the operator-Schmidt framework for bipartite dynamics~\cite{Nielsen2003}, provide the finite-layer capacity baseline needed to isolate the new question of repeated saturation.

For binary cross-adjacency matrix $\mathsf A_{AB}$, the standard cut-rank and operator-Schmidt relations give~\cite{Hein2004,Fattal2004,Nielsen2003}
\begin{align}
r_\partial^{\rm Cl}
&\equiv \operatorname{rank}_{\mathbb F_2}(\mathsf A_{AB}),
\label{eq:clifford_cut_rank}\\
C_\partial^{\rm layer}
&\equiv \log_2\operatorname{OSR}[U_\partial(t_0)]
=r_\partial^{\rm Cl}.
\label{eq:clifford_layer_capacity}
\end{align}
We measure bipartite entanglement in ebits, defining $E_A\equiv S_A/\ln2$, so that a maximally entangled qubit pair carries one ebit. For stabilizer trajectories the same standard structure implies~\cite{Fattal2004,Nielsen2003}
\begin{align}
E_A^{(\ell+1)}-E_A^{(\ell)}&\leq r_\partial^{\rm Cl},
\label{eq:clifford_layer_increment}\\
L&\geq \left\lceil\frac{K}{r_\partial^{\rm Cl}}\right\rceil
\label{eq:clifford_layer_lower_bound}
\end{align}
to reach $K$ ebits from a product stabilizer state. These standard relations are used only as a rigorous capacity-counting baseline. Appendix~\ref{app:clifford_cut_rank} summarizes the reduction.

We fix
\begin{equation}
\begin{aligned}
A&=\{1,2,3,4\},\qquad B=\{5,6,7,8\},\\
E_\partial&=\{(4,5),(3,6)\}.
\end{aligned}
\label{eq:fixed_interface_sets}
\end{equation}
so that every architecture has the same cross-cut Hamiltonian
\begin{equation}
H_\partial=J(Z_4Z_5+Z_3Z_6).
\label{eq:fixed_interface}
\end{equation}
Only edges internal to $A$ and $B$ are varied. Because all Ising terms commute, the internal evolution factors are local with respect to the bipartition and do not change the cross-cut entanglement. The two independent crossing bonds have $r_\partial^{\rm Cl}=2$. The standard cut-rank identity~\cite{Hein2004,Fattal2004} therefore gives, for every architecture in the family,
\begin{equation}
C_\partial^{\rm layer}=2,
\qquad
E_1\equiv \frac{S_A[U_G(t_0)\ket{+}^{\otimes8}]}{\ln2}=2\text{ebits}.
\label{eq:fixed_interface_baseline}
\end{equation}
Motivated by the $\pi/2$ single-qubit kicks used in kicked-Ising entanglement protocols~\cite{Mishra2015}, we define the site-resolved binary on-site kick alphabet
\begin{equation}
\mathcal C_{\rm bin}
=\left\{\prod_{j=1}^{8}[R_x^{(j)}(\pi/2)]^{s_j}:s_j\in\{0,1\}\right\}.
\label{eq:binary_x_pulse}
\end{equation}
Here a kick class $\mathcal C$ means a specified set of instantaneous products of on-site $X$ rotations inserted between successive physical Ising layers $U_G(t_0)$, with $U_G(t)$ and $t_0$ defined in Eqs.~\eqref{eq:uniform_ising_evolution} and~\eqref{eq:clifford_time}. The corresponding $L$-layer propagator class is
\begin{equation}
\mathcal K_{\mathcal C}^{[L]}(G)
\equiv
\left\{
U_G(t_0)P_{L-1}\cdots P_1U_G(t_0):
P_r\in\mathcal C
\right\}.
\label{eq:fixed_layer_propagator_class}
\end{equation}
Thus $\mathcal C$ specifies the allowed kick at a single intermediate step, whereas $\mathcal K_{\mathcal C}^{[L]}(G)$ is the full set of $L$-layer protocols constructed from that alphabet. We write $\mathcal K_{\rm bin}^{[L]}\equiv\mathcal K_{\mathcal C_{\rm bin}}^{[L]}$. For such a class, define the optimized entanglement after $L$ physical Ising layers by
\begin{align}
E_{L,\max}^{(\mathcal C)}(G)
&\equiv \max_{U\in\mathcal K_{\mathcal C}^{[L]}(G)}
\frac{S_A[U\ket{+}^{\otimes8}]}{\ln2},\nonumber\\
&=\max_{P_1,\ldots,P_{L-1}\in\mathcal C}
\frac{S_A(\psi_L)}{\ln2},\nonumber\\
\ket{\psi_L}
&=U_G(t_0)P_{L-1}\cdots P_1U_G(t_0)\ket{+}^{\otimes8},
\label{eq:fixed_layer_entanglement}
\end{align}
and the minimum saturation depth for a target of $K$ ebits by
\begin{equation}
L_*^{(\mathcal C)}(G,K)
\equiv\min\left\{L:E_{L,\max}^{(\mathcal C)}(G)\geq K\right\}.
\label{eq:saturation_depth_definition}
\end{equation}
We abbreviate $E_{L,\max}^{(\mathcal C_{\rm bin})}$ as $E_{L,\max}^{\rm bin}$. For the equal $4|4$ bipartition considered below, saturation means reaching the maximal value $K=4$ ebits. Varying the 12 internal edges generates $2^{12}=4096$ labeled architectures with the same physical interface, layer capacity, and first-layer entanglement, yet exhaustive two-layer optimization already gives different values of $E_{2,\max}^{\rm bin}$. Thus fixed interface capacity does not determine repeated utilization. To isolate architecture from connectedness and edge budget, we restrict to connected halves with six internal edges, yielding the $16\times16=256$ tree--tree family shown in Fig.~\ref{fig:rooted_replenishment_map}(a). The enumeration details and the larger-family counts are given in Appendix~\ref{app:fixed_interface_ensemble}. Exhaustive optimization of this controlled family gives the exact depth split
\begin{equation}
L_*^{(\mathcal C_{\rm bin})}(G,4)
=\begin{cases}
2,&64\ \text{graphs},\\
3,&192\ \text{graphs}.
\end{cases}
\label{eq:tree_tree_depth_counts}
\end{equation}
Every two-layer failure reaches four ebits in three binary layers.

The central architectural result is that this depth split is exactly classified by \emph{rooted} internal topology. The distinguished interface sets are $\partial A=\{3,4\}$ and $\partial B=\{5,6\}$. For $X=A,B$, define
\begin{equation}
d_X^{\max}
=\max_{v\in X}\min_{b\in\partial X}d_{G[X]}(v,b),
\label{eq:interface_domination_distance}
\end{equation}
which is the largest distance of any vertex on side $X$ from its nearest interface vertex. The interface set is dominating when every vertex lies within graph distance one of it. In the present family each side contains noninterface vertices, so this condition is equivalent to $d_X^{\max}=1$. Figure~\ref{fig:rooted_replenishment_map}(c) illustrates the distinction with two full $N=8$ tree--tree architectures. In the upper example, the numbered-vertex distances to the nearest interface vertex have maxima $d_A^{\max}=d_B^{\max}=1$. In the lower example, vertex 1 lies two internal edges from $\partial A$, giving $d_A^{\max}=2$ while $d_B^{\max}=1$, even though the internal graph on $A$ is still a $P_4$. Thus the relevant variable is the internal topology together with its placement relative to the fixed interface.

We next compare these rooted descriptors with the depths $L_*^{(\mathcal C_{\rm bin})}$ determined by the exhaustive optimization above. Across all 256 architectures, the data obey
\begin{equation}
L_*^{(\mathcal C_{\rm bin})}(G,4)
=\begin{cases}
2,&\substack{G[A]\cong G[B]\cong P_4,\\ d_A^{\max}=d_B^{\max}=1},\\
3,&\text{otherwise},
\end{cases}
\label{eq:tree_tree_topology_rule}
\end{equation}
with no exceptions. Equation~\eqref{eq:tree_tree_topology_rule} is therefore an exact classification of this finite family, not an analytic formula derived from $d_X^{\max}$ alone. In particular, $d_A^{\max}=d_B^{\max}=1$ is not sufficient if either internal tree is $K_{1,3}$. Among the 144 $P_4/P_4$ architectures, however, the distance criterion is decisive: the 64 rooted placements for which both interface sets dominate have $L_*=2$, while the remaining 80 require three layers. Figure~\ref{fig:rooted_replenishment_map}(b) shows the corresponding exhaustive depth map.

\begin{figure*}[t]
\centering
\includegraphics[width=\textwidth]{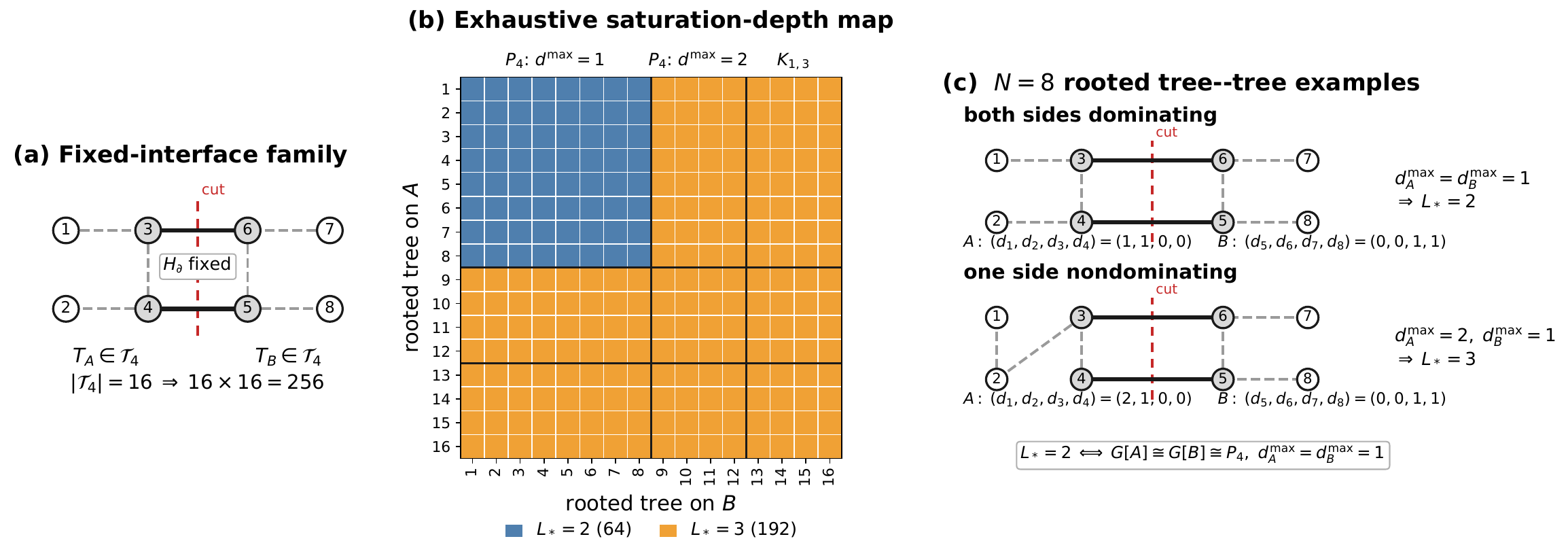}
\caption{\textbf{Rooted architecture exactly classifies interface replenishment in the controlled $N=8$ tree--tree family.} (a) Representative member of the fixed-interface family. Thick solid bonds are the two fixed cross-cut Ising couplings, dashed bonds are the variable internal tree couplings, and the red dashed line marks the bipartition. All members share the baseline in Eq.~\eqref{eq:fixed_interface_baseline}. (b) Exhaustive minimum-depth map over all $16\times16=256$ labeled tree--tree architectures under $\mathcal C_{\rm bin}$. Of these, 64 have $L_*=2$ and 192 have $L_*=3$. (c) Representative $N=8$ examples of dominating and nondominating rooted architectures. Gray vertices mark the interface sets, and the displayed tuples give the numbered-vertex distances to the nearest interface vertex. The complete classification is given by Eq.~\eqref{eq:tree_tree_topology_rule}.}
\label{fig:rooted_replenishment_map}
\end{figure*}

As a robustness check, enlarge the allowed local kicks to
\begin{equation}
\begin{aligned}
\mathcal C_{X_4}
&=\left\{\prod_{j=1}^{8}R_x^{(j)}(n_j\pi/2):
 n_j\in\{0,1,2,3\}\right\},\\
\mathcal C_{\rm bin}&\subset\mathcal C_{X_4}.
\end{aligned}
\label{eq:x4_pulse_class}
\end{equation}
We correspondingly write $\mathcal K_{X_4}^{[L]}\equiv\mathcal K_{\mathcal C_{X_4}}^{[L]}$ for the $L$-layer protocol class generated by these quarter-turn kicks. A complete two-layer search gives the same optimum graph by graph, so the depth split and rooted-topology rule above are unchanged in $\mathcal C_{X_4}$. Hence, after interface capacity, first-layer entanglement, connectedness, and internal edge budget are fixed, the minimum saturation depth remains architecture dependent and is exactly classified, within this controlled finite family, by rooted topology. This identifies architecture-dependent replenishment as a resource distinct from interface capacity.

\section{Replenishment Beyond Fixed-Layer Control}
\label{sec:beyond_fixed_layer}

The exact depth $L_*$ above is defined for restricted fixed-duration layers. We now increase the temporal resolution of the same on-site $x$-control resource to separate a discrete layer stall from the underlying continuous replenishment dynamics.

\subsection{Fast-X hierarchy and continuous control}
\label{sec:fast_x_hierarchy}

To compare the fixed-layer protocols with higher-resolution control under the same interaction-time resource, we introduce a nested fast-$X$ kicked-control hierarchy. This construction follows the standard fast-local and locally assisted control setting, in which local rotations are treated as arbitrarily fast while the finite-strength interaction time is the costly resource~\cite{Khaneja2001,Bennett2003,Malvetti2024}. For a fixed graph $G$, total interaction time $T$, and $M$ Ising-evolution segments, define the $M$-segment fast-$X$ class
\begin{equation}
\begin{split}
\mathcal K_X^{[M]}(T;G)=\Big\{&U_G(\tau_M)P_{M-1}\cdots P_1U_G(\tau_1):\\
&P_r=\prod_jR_x^{(j)}(\theta_{j,r}),\quad \theta_{j,r}\in\mathbb R,\\
&\tau_r\geq0,\quad \sum_{r=1}^{M}\tau_r=T\Big\}.
\end{split}
\label{eq:fast_x_kick_class}
\end{equation}
Here $M$ counts interaction segments. Combining Eq.~\eqref{eq:fixed_layer_propagator_class} with $\mathcal C_{\rm bin}\subset\mathcal C_{X_4}$ gives, at total interaction time $Lt_0$,
\begin{equation}
\mathcal K_{\rm bin}^{[L]}(G)
\subset
\mathcal K_{X_4}^{[L]}(G)
\subset
\mathcal K_X^{[L]}(Lt_0;G).
\label{eq:kick_hierarchy_inclusion}
\end{equation}
The first inclusion enlarges the allowed discrete kick alphabet, while the second additionally allows arbitrary $X$-rotation angles and variable Ising-segment durations at fixed total interaction time. The chain protocols of Sec.~\ref{sec:optimal_routing} are structured members of this hierarchy. For numerical convergence tests we use
\begin{equation}
E_{X,M}^{\rm kick}(T;G)
\equiv
\sup_{U\in\mathcal K_X^{[M]}(T;G)}
\frac{S_A[U\ket{+}^{\otimes N}]}{\ln2}.
\label{eq:finite_fast_x_envelope}
\end{equation}
Increasing $M$ probes the arbitrary-resolution kicked envelope. No finite $M$ is identified with its exact limit.

The continuous VEEF control uses the same local axis,
\begin{equation}
H_{\rm VEEF}(t;G)
=H_G+\sum_{j=1}^{8}h_j^x(t)X_j,
\label{eq:veef_xonly}
\end{equation}
with no independently controlled $Y$ or $Z$ fields~\cite{Lu2021,Lu2024}. A fine Strang splitting~\cite{Strang1968,ChildsTrotter2021} alternates Ising evolution with arbitrary local $X$ rotations, so finite-step VEEF trajectories lie within the same high-resolution fast-$X$ structure up to the discretization error. The fine-graining strategy increases temporal resolution while maintaining high precision and numerical stability in the variational field optimization~\cite{Lu2021}. We therefore use the kicked hierarchy and VEEF as independent parameterizations of the same idealized $x$-only interaction-time resource. The formal continuous-control envelope is
\begin{equation}
E_x^\star(T;G)
\equiv
\sup_{\{h_j^x(t)\}}
\frac{S_A(T)}{\ln2}.
\label{eq:xonly_control_envelope}
\end{equation}
Neither a finite-$M$ kicked optimization nor a VEEF trajectory is taken as an analytic upper bound on $E_x^\star$. Throughout this section, the optimized resource is \emph{interaction time} under the idealization of arbitrarily fast on-site rotations. Peak control amplitude, pulse bandwidth, and wall-clock duration define separate resource dimensions~\cite{Hirose2018,Koch2022}. The VEEF Trotter resolution is a numerical discretization parameter rather than a physical layer count, so the kicked and VEEF calculations compare interaction-time performance under the shared $x$-only control resource. Full optimization and reproducibility details are given in Appendix~\ref{app:veef_numerics}.

\subsection{Two-channel stall relaxation}

The two-channel graph
\begin{equation}
H_{\rm 2ch}
=J\sum_{j=1}^{7}Z_jZ_{j+1}+JZ_3Z_6
\label{eq:H2ch}
\end{equation}
is a concrete $L_*=3$ member of the tree--tree family with the same two-channel interface as Eq.~\eqref{eq:fixed_interface}. The cut-rank baseline is two layers for a four-ebit target, yet exhaustive searches give
\begin{equation}
L_*^{(\mathcal C_{\rm bin})}(G_{\rm 2ch},4)
=L_*^{(\mathcal C_{X_4})}(G_{\rm 2ch},4)=3.
\label{eq:two_channel_stall}
\end{equation}
The explicit three-layer binary construction is given in Appendix~\ref{app:routing_obstruction}.

To determine whether this discrete three-layer stall persists once the temporal resolution is increased, we compare the interaction time required to reach near-maximal entanglement under the fast-$X$ hierarchy and VEEF. For reference, at $T=2t_0$ the restricted fixed-layer classes have the exact ceiling
\begin{equation}
E_{2,\max}^{\rm bin}=E_{2,\max}^{X_4}=2\text{ebits}.
\end{equation}
For the near-maximal threshold $E_A\geq3.99$ ebits, define
\begin{equation}
T_{X,M}^{(0.01)}
\equiv
\inf\{T:E_{X,M}^{\rm kick}(T;G_{\rm 2ch})\geq3.99\}.
\label{eq:fast_x_threshold_time}
\end{equation}
The independent kicked and VEEF searches are summarized in Table~\ref{tab:fast_x_crossing}. The $M=5,7,9$ kicked optimizations converge to threshold times between $2.8765t_0$ and $2.8807t_0$, while VEEF gives $T_{\rm VEEF}^{(0.01)}\simeq2.8786t_0$. The agreement of two independent parameterizations identifies $T\simeq2.88t_0$ as the numerically resolved fast-$X$ optimum for this architecture at the present resolutions. This is already below the exact restricted-layer time $3t_0$, showing that the three-layer stall reflects finite temporal resolution rather than a fundamental limit of the broader $x$-only interaction-time resource. Further resolution can sharpen the continuum value.

\begin{table}[tb]
\centering
\caption{Near-maximal target times for the two-channel graph with threshold $E_A\geq3.99$ ebits. The kicked entries use variable segment durations, VEEF is optimized independently, and all reported crossing times are obtained by linear interpolation between the nearest fine-scan points bracketing the threshold.}
\label{tab:fast_x_crossing}
\begin{tabular}{lcc}
Control & Resolution & $T^{(0.01)}/t_0$ \\
\hline
fast-$X$ kicks & $M=5$ & 2.880684 \\
fast-$X$ kicks & $M=7$ & 2.878436 \\
fast-$X$ kicks & $M=9$ & 2.876500 \\
VEEF & continuous field & 2.878633
\end{tabular}
\end{table}

\subsection{Dynamical identification of rooted replenishment classes}
\label{sec:veef_fingerprint}

The two-channel calibration above establishes that the discrete layer depth can be softened by higher-resolution $x$-only control. We next ask whether the rooted replenishment classes of Eq.~\eqref{eq:tree_tree_topology_rule} remain distinguishable in the optimized dynamics after that restriction is relaxed. We therefore apply the same VEEF procedure to all 21 rooted orbits obtained from the 256 tree--tree graphs after the $A\leftrightarrow B$ symmetry reduction. The optimizer is not supplied with $L_*$, a rooted-topology class label, or a graph-specific analytic pulse sequence. All orbits use the same $x$-only parameterization, optimizer settings, regularization weight $\lambda=0.05$, and eight independent starts. Let $E_{\rm VEEF}(T;G)$ denote the best entanglement found by this common numerical procedure.

We use $T=2t_0$ as a common diagnostic time rather than as an optimized arrival time. At this time all six $L_*=2$ orbits are essentially saturated, with $E_{\rm VEEF}\geq3.999530$, whereas all 15 $L_*=3$ orbits remain below $3.713595$ ebits. A pre-specified single-time threshold $E_c=3.70$ ebits identifies $18/21$ orbits and leaves the boundary cases O08, O16, and O17 unresolved. The full orbit-resolved convergence and stability checks are given in Appendix~\ref{app:veef_validation}.

\begin{figure*}[t]
\centering
\includegraphics[width=1\textwidth]{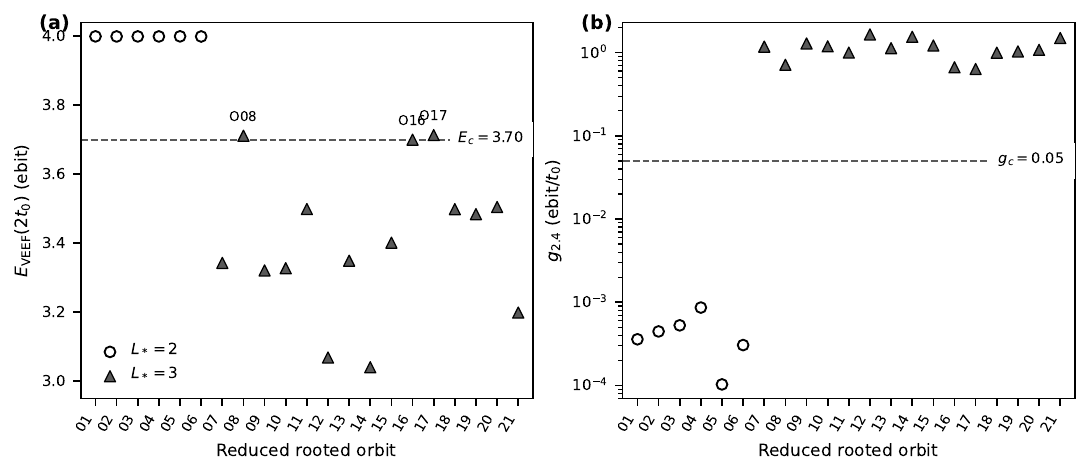}
\caption{\textbf{Time-resolved VEEF identifies the rooted replenishment partition across the complete symmetry-reduced family.} (a) Best VEEF entanglement at the common diagnostic time $T=2t_0$ for all 21 reduced rooted orbits. The pre-specified threshold $E_c=3.70$ leaves the three $L_*=3$ boundary orbits O08, O16, and O17 unresolved. (b) The frozen two-time growth statistic $g_{2.4}$ separates all six $L_*=2$ orbits from all 15 $L_*=3$ orbits using the fixed threshold $g_c=0.05$ ebit/$t_0$. The logarithmic scale shows the nearly three-order-of-magnitude separation between already saturated and still-replenishing architectures.}
\label{fig:veef_dynamical_recovery}
\end{figure*}

These unresolved single-time cases motivate a dynamical observable. Before the initial second-time calculations we fixed $T_2=2.4t_0$ and the finite-difference average entanglement-growth rate of the optimized VEEF envelope,
\begin{equation}
g_{2.4}(G)
\equiv
\frac{E_{\rm VEEF}(2.4t_0;G)-E_{\rm VEEF}(2t_0;G)}{0.4t_0}.
\label{eq:veef_two_time_growth}
\end{equation}
We also fixed the threshold $g_c=0.05$ ebit/$t_0$, assigning $L_*=3$ when $g_{2.4}>g_c$. After the nine-orbit boundary diagnostic, the same statistic, threshold, optimizer settings, and random seed were frozen and applied to the remaining 12 $L_*=3$ orbits. Over the complete 21-orbit reduced family,
\begin{align}
1.03\times10^{-4}
&\leq g_{2.4}\leq8.64\times10^{-4}
&&\text{ebit}/t_0,\qquad L_*=2,\nonumber\\
0.636097
&\leq g_{2.4}\leq1.655758
&&\text{ebit}/t_0,\qquad L_*=3.
\label{eq:veef_two_time_ranges}
\end{align}
The two classes are therefore separated by nearly three orders of magnitude in growth rate, and the frozen rule identifies all $21/21$ rooted orbits without threshold adjustment. This complete separation is the continuous-dynamical counterpart of the fixed-layer rooted classification in Fig.~\ref{fig:rooted_replenishment_map}: architectures in the $L_*=2$ class have effectively exhausted the available entangling opportunity by $2t_0$, while every $L_*=3$ architecture remains actively replenishing the interface over the next $0.4t_0$. Figure~\ref{fig:veef_dynamical_recovery} makes this connection explicit. Panel (a) shows that a single-time snapshot leaves the three boundary cases unresolved under the pre-specified threshold, whereas panel (b) shows that the time-resolved growth cleanly identifies the full rooted partition. The rooted classification is therefore not only a property of a restricted layer count but a directly resolvable signature of optimized continuous dynamics.

A representative $\lambda$ scan on O06, O17, and O20 changes the optimized scores by less than $0.05$ ebit and preserves their ordering, supporting stability to moderate regularization changes. Numerical details are given in Appendix~\ref{app:veef_validation}. Together with Fig.~\ref{fig:veef_dynamical_recovery}, these results complete the paper's progression from interface capacity to rooted replenishment and finally to dynamical identification.

\section{Discussion and Perspective}
\label{sec:discussion}

The main conceptual separation is between the entangling flux available at a fixed physical interface and the ability of the surrounding architecture to sustain its use. Entangling capacities, entropy-rate bounds, and fast-local control limits provide the broader setting~\cite{Dur2001,Bennett2003,Childs2003,Bravyi2007,VanAcoleyen2013,Malvetti2024}. In fermionic Gaussian dynamics, $\|K_{AB}\|_*/2$ gives an exact bound on the collective speed of the canonical entanglement angles, while the utilization $u(t)$ distinguishes available flux from realized flux. The Majorana-rematching chain constructions saturate this capacity and convert the bound into exact minimum interaction times under the stated control model, with kicked-Ising and time-optimal spin-control protocols providing relevant antecedents~\cite{Fisher2009,Mishra2015}. This interaction-time optimality is complementary to recent optimal preparation of fermionic Gaussian states formulated in terms of matchgate gate count and circuit depth~\cite{Langer2026}.

Holding the cross-cut interaction, first-layer entanglement, connectedness, and internal edge budget fixed isolates replenishment as an architectural resource. Standard cut-rank and stabilizer relations provide the fixed-layer baseline~\cite{Hein2004,Fattal2004,Nielsen2003}, while the exhaustive $N=8$ tree--tree family shows that internal tree type and rooted interface placement jointly determine whether the fixed interface can be saturated in two or three restricted layers. The effect is therefore not a change in interface capacity but a change in how efficiently fresh degrees of freedom reach and reuse that interface. This viewpoint is complementary to graph-state preparation complexity and circuit-structure studies~\cite{Cabello2011,Kumabe2026,Rao2026} and to routing and bottleneck bounds for quantum information transport~\cite{Eldredge2020,Bapat2023,Devulapalli2026}.

Higher-resolution $x$-only control promotes this finite-depth distinction into a continuous dynamical one. The fast-$X$ hierarchy lies in the standard fast-local interaction-time setting~\cite{Khaneja2001,Caneva2009,Malvetti2024}, while VEEF supplies an independent continuous-field parameterization with fine-grained variational optimization~\cite{Lu2021,Lu2024}. In the calibrated two-channel architecture, VEEF reaches the numerically resolved fast-$X$ optimum, and across all 21 symmetry-reduced rooted orbits its time-resolved growth identifies exactly the same $L_*=2/3$ replenishment partition found by exhaustive fixed-layer optimization. The resulting physical picture is capacity $\rightarrow$ rooted replenishment $\rightarrow$ dynamical identification. A central open problem is to turn this architecture-dependent replenishment structure into a general non-Gaussian analytic bound on repeated interface utilization beyond the finite family studied here.

\begin{acknowledgments}
This work was supported in part by the Strategic Priority Research Program of the Chinese Academy of Sciences (Grant No.~XDB1270000) and the robotic AI-Scientist platform of the Chinese Academy of Sciences.
\end{acknowledgments}

\section*{Data Availability}
Analytical derivations are contained in the article. Numerical data and code supporting the figures will be made available in a public repository upon publication.

\appendix

\section{Proof of the Gaussian Angle-Flow Bound}
\label{app:gaussian_angle_flow_proof}

For completeness, we give the full derivation of Theorem~\ref{thm:gaussian_angle_flow}. Define the covariance matrix
\begin{equation}
\Gamma_{pq}
=\frac{\ii}{2}\langle[\gamma_p,\gamma_q]\rangle .
\end{equation}
With the Hamiltonian convention of Eq.~\eqref{eq:majorana_hamiltonian}, the Majorana operators obey $\dot\gamma=K\gamma$, and hence~\cite{BravyiFermionic2005}
\begin{equation}
\dot\Gamma=[K,\Gamma].
\label{eq:covariance_flow}
\end{equation}
For a pure Gaussian state, $\Gamma^2=-I$. At a regular time, choose local orthogonal frames on $A$ and $B$ that put the $k$th entangled pair into canonical form. On the corresponding two-dimensional Majorana plane of $A$ and its paired plane in $B$,
\begin{equation}
\Gamma_{A,k}=c_kJ_2,
\qquad
\Gamma_{AB,k}=s_kR_k,
\label{eq:canonical_covariance_blocks}
\end{equation}
where $c_k=\cos(2\theta_k)$, $s_k=\sin(2\theta_k)$, $J_2=\begin{psmallmatrix}0&1\\-1&0\end{psmallmatrix}$, and $R_k$ is orthogonal. The second relation follows from purity because the $k$th diagonal block of $\Gamma^2=-I$ gives $\Gamma_{AB,k}\Gamma_{AB,k}^{T}=s_k^2I_2$.

The $AA$ block of Eq.~\eqref{eq:covariance_flow} is
\begin{equation}
\dot\Gamma_A
=[K_A,\Gamma_A]
-K_{AB}\Gamma_{AB}^{T}
+\Gamma_{AB}K_{AB}^{T}.
\label{eq:covariance_A_block}
\end{equation}
The local commutator changes the canonical basis but not the canonical eigenvalue $c_k$. We may therefore evaluate its derivative in a co-moving local frame. Let $A_k$ denote the $2\times2$ block of the physical cross-boundary generator $K_{AB}$ between these paired canonical planes. Since the canonical form has no other cross-covariance block in the $k$th row, Eq.~\eqref{eq:covariance_A_block} gives
\begin{equation}
\left.\dot\Gamma_{A,k}\right|_{\partial}
=-A_k(s_kR_k)^T+s_kR_kA_k^T.
\end{equation}
Because $c_k=-\tfrac12\Tr(J_2\Gamma_{A,k})$, projection onto $J_2$ yields
\begin{align}
\dot c_k
&=-\frac12\Tr\!\left[J_2\left.\dot\Gamma_{A,k}\right|_{\partial}\right]\\
&=-s_k\Tr(A_k^TJ_2R_k).
\label{eq:ck_flow}
\end{align}
For an interior regular pair, $\dot c_k=-2s_k\dot\theta_k$, and therefore
\begin{equation}
\dot\theta_k
=\frac12\Tr(A_k^TJ_2R_k).
\label{eq:theta_block_flow}
\end{equation}
At regular endpoints the same bound follows by continuity whenever the canonical-angle derivative exists, which is sufficient for the almost-everywhere statement in Theorem~\ref{thm:gaussian_angle_flow}.

Now take the singular-value decomposition
\begin{equation}
K_{AB}=2\sum_{\alpha=1}^{r}\lambda_\alpha
u^{(\alpha)}v^{(\alpha)T},
\qquad \lambda_\alpha\geq0,
\label{eq:Kab_svd_appendix}
\end{equation}
where the left and right singular vectors are normalized. Denote their projections onto the $k$th canonical planes by $u_k^{(\alpha)}$ and $v_k^{(\alpha)}$. Substitution into Eq.~\eqref{eq:theta_block_flow} gives
\begin{equation}
\dot\theta_k
=\sum_{\alpha=1}^{r}\lambda_\alpha a_{k\alpha},
\qquad
a_{k\alpha}
=u_k^{(\alpha)T}J_2R_kv_k^{(\alpha)}.
\label{eq:theta_singular_channel_flow}
\end{equation}
Hence
\begin{align}
|a_{k\alpha}|
&\leq\|u_k^{(\alpha)}\|_2\,\|v_k^{(\alpha)}\|_2,\\
\sum_{k=1}^{m}|a_{k\alpha}|
&\leq
\left(\sum_k\|u_k^{(\alpha)}\|_2^2\right)^{1/2}
\left(\sum_k\|v_k^{(\alpha)}\|_2^2\right)^{1/2}
\leq1,
\end{align}
where the last step uses normalization of the singular vectors. Components in unpaired local Majorana planes only reduce the two sums. Therefore
\begin{equation}
\sum_k|\dot\theta_k|
\leq\sum_\alpha\lambda_\alpha
=\frac12\|K_{AB}\|_*.
\end{equation}
Finally, local quadratic frame transformations act by independent orthogonal rotations on the left and right of $K_{AB}$ and leave its singular values, and hence the bound, invariant.

\section{Derivation of the Standard Cut-Rank Layer-Capacity Bound}
\label{app:clifford_cut_rank}

For completeness, we collect the standard graph-state and stabilizer steps underlying Eqs.~\eqref{eq:clifford_layer_capacity}--\eqref{eq:clifford_layer_lower_bound}~\cite{Hein2004,Fattal2004,Nielsen2003}. Because all $ZZ$ terms commute, one full layer factorizes as
\begin{equation}
U_G(t_0)=U_AU_BU_\partial,
\end{equation}
where $U_A$ and $U_B$ contain edges internal to the two sides and are local with respect to the bipartition. For each crossing edge, $\exp[-\ii Jt_0Z_iZ_j]$ is locally Clifford equivalent to $CZ_{ij}$, so, up to local Clifford factors,
\begin{equation}
U_\partial\sim
\prod_{i\in A,j\in B}
CZ_{ij}^{(\mathsf A_{AB})_{ij}}.
\label{eq:cross_cz_network}
\end{equation}
In the computational basis this network contributes the bilinear phase
\begin{equation}
(-1)^{\mathbf x_A^T\mathsf A_{AB}\mathbf x_B}.
\end{equation}
Invertible binary basis changes on $A$ and $B$, implementable by Clifford CNOT networks local to the bipartition, reduce $\mathsf A_{AB}$ by Gaussian elimination to a matrix with an $r_\partial^{\rm Cl}\times r_\partial^{\rm Cl}$ identity block and zeros elsewhere. Hence $U_\partial$ is locally equivalent to $r_\partial^{\rm Cl}$ independent controlled-$Z$ gates. Its operator-Schmidt rank is therefore
\begin{equation}
\operatorname{OSR}(U_\partial)=2^{r_\partial^{\rm Cl}}.
\end{equation}
This is the unitary counterpart of the standard graph-state cut-rank formula~\cite{Hein2004,Fattal2004}.

If a pure state has Schmidt rank $\chi$, application of an operator with Schmidt rank $R$ produces a state of Schmidt rank at most $R\chi$~\cite{Nielsen2003}. For stabilizer states the nonzero Schmidt coefficients across a bipartition are equal~\cite{Fattal2004}, so $S_A/\ln2=\log_2\chi$. Therefore one Ising layer can increase the entanglement by at most $\log_2\operatorname{OSR}(U_\partial)=r_\partial^{\rm Cl}$ ebits. Local Clifford operations between layers preserve the bipartite entropy, proving Eq.~\eqref{eq:clifford_layer_increment} and its iteration Eq.~\eqref{eq:clifford_layer_lower_bound}.

\section{Fixed-Interface Exhaustive Enumeration}
\label{app:fixed_interface_ensemble}

The fixed-interface family in Sec.~\ref{sec:fixed_interface_ensemble} keeps $E_\partial=\{(4,5),(3,6)\}$ unchanged and independently includes or excludes each of the six possible internal edges on $A=\{1,2,3,4\}$ and the six possible internal edges on $B=\{5,6,7,8\}$. This gives $2^{12}=4096$ labeled graphs. For each graph we first evaluate $U_G(t_0)\ket{+}^{\otimes8}$ and verify $E_1=S_A/\ln2=2$. The maximum absolute numerical error over the ensemble is $4.4\times10^{-16}$. We then enumerate all $2^8$ pulse patterns in $\mathcal C_{\rm bin}$ for $U_G(t_0)P(\mathbf s)U_G(t_0)$ and record the maximum entanglement $E_{2,\max}^{\rm bin}$.

The full ensemble contains $704$, $2816$, and $576$ graphs with $E_{2,\max}^{\rm bin}=2,3,4$, respectively. A graph is labeled interface reachable when every connected component on side $A$ contains at least one of the interface vertices $\{3,4\}$ and every connected component on side $B$ contains at least one of $\{5,6\}$. There are $2304$ such graphs, with counts $80$, $1648$, and $576$ in the same three entanglement classes. Requiring both internal graphs to be connected gives $1444$ graphs, with counts $48$, $912$, and $484$.

Finally, among the side-connected graphs, fixing the total internal edge count to six forces three edges on each four-vertex side. Each side is therefore a tree, giving $16\times16=256$ labeled tree--tree networks. Their two-layer counts are $22$, $170$, and $64$ for $E_{2,\max}^{\rm bin}=2,3,4$. The enumeration optimizes the allowed pulse pattern independently for every graph. The control class is identical across the ensemble, but the maximizing binary pattern need not be.

For each of the 192 graphs with $E_{2,\max}^{\rm bin}<4$, we next enumerate all $2^{16}=65536$ pairs of intermediate binary pulse patterns in a three-layer protocol. Every graph reaches four ebits, establishing $L_*^{(\mathcal C_{\rm bin})}=3$ for all two-layer failures. Independently, for every tree--tree graph we enumerate all $4^8=65536$ local quarter-turn patterns in $\mathcal C_{X_4}$ at two layers. The resulting optimum equals the binary optimum graph by graph, $E_{2,\max}^{X_4}=E_{2,\max}^{\rm bin}$. Since $\mathcal C_{\rm bin}\subset\mathcal C_{X_4}$, the binary three-layer constructions then establish the identical depth split $L_*^{(\mathcal C_{X_4})}=2$ for 64 graphs and $3$ for 192.

To classify the geometry, each four-vertex tree is labeled by its unrooted type $P_4$ or $K_{1,3}$, and for each side we evaluate $d_X^{\max}$ from Eq.~\eqref{eq:interface_domination_distance}. The 256 rows obey Eq.~\eqref{eq:tree_tree_topology_rule} without exception. Any $K_{1,3}$ side implies $L_*=3$. Among the 144 $P_4/P_4$ architectures, the 64 for which $d_A^{\max}=d_B^{\max}=1$ have $L_*=2$, while the remaining 80 have $L_*=3$. This is a complete enumeration of the stated finite family. No analytic generalization beyond it is assumed.

\section{Majorana-Rematching Protocol and Capacity Saturation}
\label{app:majorana_matching}

We give a compact constructive proof of the Majorana-rematching kick protocol $\Pi_{\rm MR}$ and the saturation statements in Sec.~\ref{sec:optimal_routing}. Using the Jordan--Wigner Majoranas defined in Eq.~\eqref{eq:jw_majoranas_main}~\cite{JordanWigner1928,LiebSchultzMattis1961}, one has
\begin{equation}
X_j=\ii a_jb_j,
\qquad
Z_jZ_{j+1}=\ii b_ja_{j+1}.
\label{eq:majorana_ising_map}
\end{equation}
Up to signs, which do not affect the entanglement, evolution for $t_0=\pi/(4|J|)$ swaps the two Majoranas on every Ising bond,
\begin{equation}
b_j\longleftrightarrow a_{j+1},
\end{equation}
while an $R_x^{(j)}(\pi/2)$ pulse swaps $a_j\leftrightarrow b_j$. The $X$-polarized product state has covariance matching
\begin{equation}
\mathcal M_0=\{(a_j,b_j):j=1,\ldots,N\}.
\end{equation}
Because every operation in the kicked protocol is a signed Majorana permutation, the state remains a perfect-matching Gaussian state. If $n_\times$ matching edges cross the bipartition, then
\begin{equation}
S_A=\frac{n_\times}{2}\ln2.
\label{eq:matching_entropy}
\end{equation}

For the open chain, apply after the $r$th physical Ising layer the nested on-site swap
\begin{equation}
P_r=\prod_{j=r+1}^{N-r}R_x^{(j)}(\pi/2).
\end{equation}
Tracking the endpoints of the matching under the alternating bond and on-site swaps gives, after the $r$th free-evolution layer,
\begin{equation}
n_\times^{\rm OBC}(r)=2r,
\qquad 1\leq r\leq m.
\end{equation}
Eq.~\eqref{eq:matching_entropy} therefore gives $S_A=r\ln2$ after $r$ layers and $S_A=m\ln2$ at $T=mt_0$. This achieves the lower bound in Eq.~\eqref{eq:TOBC}.

For the periodic chain in a fixed parity sector, the Jordan--Wigner boundary bond acts, again up to a parity-fixed sign, as the additional Majorana swap $b_N\leftrightarrow a_1$. Between physical Ising layers apply the global on-site swap
\begin{equation}
P=\prod_{j=1}^{N}R_x^{(j)}(\pi/2).
\end{equation}
The matching then has
\begin{equation}
n_\times^{\rm PBC}(r)=4r
\end{equation}
after $r$ physical Ising layers, until the smaller subsystem is exhausted. Thus each physical Ising layer creates two ebits. For even $m$, $r=m/2$ layers reach maximal entanglement. For odd $m$, after $(m-1)/2$ physical Ising layers only one canonical mode pair remains unsaturated. A final interval of duration $t_0/2$ completes that pair. In both cases the total interaction time is $mt_0/2$, establishing Eq.~\eqref{eq:TPBC}.

\section{Discrete Routing Obstruction in the Two-Channel Spin Network}
\label{app:routing_obstruction}

This appendix records the scope of the discrete claim used in Sec.~\ref{sec:beyond_fixed_layer}. Let
\begin{equation}
U_{\rm 2ch}(t_0)=e^{-\ii H_{\rm 2ch}t_0},
\qquad
t_0=\frac{\pi}{4|J|}.
\end{equation}
We consider the binary $X_{\pi/2}$ pulse class $\mathcal C_{\rm bin}$ of Eq.~\eqref{eq:binary_x_pulse}: free-evolution intervals of duration $t_0$ are separated by instantaneous on-site pulses, with each site receiving either $I$ or one $R_x(\pi/2)$ rotation per kick. The fixed-interface enumeration in Appendix~\ref{app:fixed_interface_ensemble} additionally establishes that the enlarged quarter-turn class $\mathcal C_{X_4}$ of Eq.~\eqref{eq:x4_pulse_class} has the same two-layer ceiling for this graph.

\begin{enumerate}[label=(\arabic*),leftmargin=*]
\item \textbf{Two physical Ising layers.} For the protocol $U_{\rm 2ch}(t_0)PU_{\rm 2ch}(t_0)$, exhaustive search over all $2^8$ products of on-site $R_x(\pi/2)$ rotations gives at most $S_A=2\ln2$. This statement is restricted to that ansatz. It is not a no-go theorem for arbitrary continuous controls.

\item \textbf{Three physical Ising layers.} The sequence
\begin{equation}
U_{\rm 2ch}(t_0)P_1U_{\rm 2ch}(t_0)P_2U_{\rm 2ch}(t_0),
\end{equation}
with, for example,
\begin{equation}
P_1=P_2=\prod_{j=2}^{7}R_x^{(j)}(\pi/2),
\end{equation}
reaches $S_A=4\ln2$. Hence the minimum layer count within $\mathcal C_{\rm bin}$ is three. Because the two-layer $\mathcal C_{X_4}$ optimum is also below four ebits and the same three-layer binary sequence belongs to $\mathcal C_{X_4}$, the minimum depth is three in both classes, giving Eq.~\eqref{eq:two_channel_stall}.
\end{enumerate}

Continuous VEEF uses the same $x$-only local control resource in a more flexible time-distributed form. A fine scan gives the $E_A\geq3.99$ threshold at $T_{\rm VEEF}^{(0.01)}\simeq2.8786t_0$. The independent arbitrary-angle fast-$X$ hierarchy gives $T_{X,M}^{(0.01)}/t_0=2.880684,2.878436,2.876500$ for $M=5,7,9$, respectively. Thus the VEEF threshold lies inside the directly resolved kicked-control convergence window and below the exact three-layer time $3t_0$. For comparison, a representative $L_*=2$ architecture crosses the same threshold below $2t_0$ at the tested finite resolution. In both depth classes, $L_*t_0$ is therefore a property of the restricted fixed-layer protocol rather than a continuous-control lower bound.

\section{Orbit-Exhaustive VEEF Validation}
\label{app:veef_validation}

\subsection{Numerical protocol and reproducibility}
\label{app:veef_numerics}

The numerical calculations use the same drift Hamiltonian $H_G$ and on-site $x$-only control resource defined in Sec.~\ref{sec:fast_x_hierarchy}. We set $\hbar=1$ and measure energies in units of $|J|$, so that $|J|=1$ and $t_0=\pi/4$; control-field amplitudes are expressed in the same energy units. For VEEF, the control field is represented by an unconstrained array $h_j^x(k)$ of shape $K\times N$, with no reflection-symmetry reduction or endpoint mask. At fixed total time $T$, the objective minimized at every fine-graining stage is
\begin{equation}
\mathcal L_{\rm VEEF}
=-E_A+\lambda\frac{1}{KN}\sum_{k=1}^{K}\sum_{j=1}^{N}|h_j^x(k)|,
\qquad \lambda=0.05,
\label{eq:veef_numerical_loss}
\end{equation}
where $E_A=S_A/\ln2$. Propagation uses the symmetric Strang step
\begin{equation}
\begin{aligned}
U_k&=R_x[\mathbf h_x(k),\tau/2]\,e^{-\ii H_G\tau}\,R_x[\mathbf h_x(k),\tau/2],\\
\tau&=T/K.
\end{aligned}
\end{equation}
where $R_x[\mathbf h_x(k),\tau/2]$ denotes the simultaneous site-resolved half-step rotations generated by the fields $h_j^x(k)$. The propagation uses complex128 arithmetic and the full $2^8\times2^8$ propagator. The target numerical step size is $\tau_{\rm target}=0.01$ in these units, with $K_{\rm target}=\operatorname{round}(T/\tau_{\rm target})$. Starting from $K=10$, the grid is doubled stage by stage until the target resolution is reached. Thus the paths are $10\to20\to40\to80\to157$ at $T=2t_0$, $10\to20\to40\to80\to188$ at $T=2.4t_0$, and $10\to20\to40\to80\to160\to236$ at $T=3t_0$. The field is initialized from a normal distribution with standard deviation $\max[0.1,0.1\sqrt{K/60}]$ and is optimized with Adam at learning rate $0.05$. A ReduceLROnPlateau scheduler monitors $E_A$ with patience $50$, reduction factor $0.5$, and minimum learning rate $10^{-4}$. Each stage allows at most $5000$ iterations and terminates after $200$ iterations without an improvement larger than $10^{-5}$ in $E_A$. Nonfinite gradient entries are set to zero. The eight-start calculations optimize all starts in batch with the same stopping rule.

All VEEF calculations use the base seed $42$. The multistart seeds are generated as $42+s\times9973$ for $s=0,\ldots,7$. In single-start fine graining, stage $r$ uses seed $42+r\times1000003$. The two-time statistic, $g_c=0.05$ ebit/$t_0$, $\lambda=0.05$, optimizer settings, eight-start protocol, and base seed were frozen before extending the initial nine-orbit diagnostic to the remaining 12 $L_*=3$ orbits. The class label $L_*$ is never supplied to the optimizer and is used only afterward for scoring.

For the fast-$X$ calculations, the local kick angles have shape $(M-1)\times N$ and are initialized independently from a normal distribution of standard deviation $0.3$. Adam is used with learning rate $0.05$ and a ReduceLROnPlateau scheduler with patience $100$, reduction factor $0.5$, and minimum learning rate $10^{-4}$. The default search uses $32$ starts and at most $3000$ iterations per start, with early termination after $300$ iterations without improvement. The highest-resolution $M=9$ trajectory calculations use $64$ starts and up to $5000$ iterations. Segment durations are either fixed at $T/M$ or optimized under the simplex constraint by a softmax parameterization. The variable-duration version is used for the crossing times in Table~\ref{tab:fast_x_crossing}. When multiple starts are optimized in a batch, the loss is the negative mean entanglement over the active starts. Coarse time scans use spacing $0.1t_0$ down to $0.5L_*t_0$, followed by a $0.05t_0$ scan in the near-saturation region. Warm starts propagate the best angles and duration parameters from one time point to the next. The convergence comparison uses $M=5,7,9$, while the broader scans use $M\in\{L_*,L_*+1,L_*+2\}$.

For both parameterizations, threshold times are extracted from the discrete scans by linear interpolation of the entanglement gap $g(T)\equiv4-E_A(T)$ between the two nearest points that bracket $g=0.01$. Writing these points as $(T_{\rm lo},g_{\rm lo})$ and $(T_{\rm hi},g_{\rm hi})$, we use
\begin{equation}
T^{(0.01)}
=T_{\rm lo}+(T_{\rm hi}-T_{\rm lo})
\frac{g_{\rm lo}-0.01}{g_{\rm lo}-g_{\rm hi}}.
\label{eq:threshold_linear_interpolation}
\end{equation}
For the $M=5,7,9$ fast-$X$ entries in Table~\ref{tab:fast_x_crossing}, the bracketing points are $T_{\rm lo}=2.85t_0$ and $T_{\rm hi}=2.90t_0$; the VEEF crossing is interpolated between $2.86t_0$ and $2.88t_0$. These interpolations give $T_{X,M}^{(0.01)}/t_0=2.880684,2.878436,2.876500$ for $M=5,7,9$, respectively, and $T_{\rm VEEF}^{(0.01)}/t_0=2.878633$.

These settings make the VEEF and fast-$X$ calculations independently reproducible at the resolutions reported here. The eight-start statistics quantify local-optimization variability at the diagnostic time, the triggered 16-start reruns test the least stable single-time cases, and the representative $\lambda$ scan tests sensitivity to moderate regularization changes. Together with the independent $M=5,7,9$ convergence in Table~\ref{tab:fast_x_crossing}, these checks support the resolved-optimum and dynamical-identification statements made in Sec.~\ref{sec:beyond_fixed_layer}.

\subsection{Orbit-resolved validation}

\begin{table*}[tb]
\centering
\caption{Single-time class-label-blind VEEF results at $T=2t_0$ for all 21 reduced rooted orbits. $E_{\rm best}$ is the best of eight starts and $E_{\rm mean}$ and $\sigma$ are the corresponding multi-start mean and standard deviation.}
\label{tab:veef_all_orbits}
\begin{tabular}{ccccc@{\qquad}ccccc}
Orbit & $L_*$ & $E_{\rm best}$ & $E_{\rm mean}$ & $\sigma$ & Orbit & $L_*$ & $E_{\rm best}$ & $E_{\rm mean}$ & $\sigma$ \\
\hline
O01 & 2 & 3.999792 & 3.999674 & $8.2\times10^{-5}$ & O12 & 3 & 3.068562 & 3.008405 & 0.0227 \\
O02 & 2 & 3.999576 & 3.999556 & $2.1\times10^{-5}$ & O13 & 3 & 3.349273 & 3.295235 & 0.0355 \\
O03 & 2 & 3.999629 & 3.999607 & $1.7\times10^{-5}$ & O14 & 3 & 3.040255 & 3.008264 & 0.0150 \\
O04 & 2 & 3.999595 & 3.999559 & $2.7\times10^{-5}$ & O15 & 3 & 3.401533 & 3.355122 & 0.0440 \\
O05 & 2 & 3.999904 & 3.999778 & $9.3\times10^{-5}$ & O16 & 3 & 3.700315 & 3.683020 & 0.0066 \\
O06 & 2 & 3.999530 & 3.999523 & $4.4\times10^{-6}$ & O17 & 3 & 3.713595 & 3.690188 & 0.0124 \\
O07 & 3 & 3.342932 & 3.293193 & 0.0391 & O18 & 3 & 3.498844 & 3.490436 & 0.0065 \\
O08 & 3 & 3.711661 & 3.687845 & 0.0105 & O19 & 3 & 3.484350 & 3.423270 & 0.0280 \\
O09 & 3 & 3.321077 & 3.276872 & 0.0465 & O20 & 3 & 3.504916 & 3.451386 & 0.0633 \\
O10 & 3 & 3.327647 & 3.304055 & 0.0224 & O21 & 3 & 3.198752 & 3.071151 & 0.0616 \\
O11 & 3 & 3.498990 & 3.489881 & 0.0065 & & & & & \\
\end{tabular}
\end{table*}

\begin{table}[tb]
\centering
\caption{Sixteen-start refinement of the seven orbits selected by the pre-specified convergence diagnostics.}
\label{tab:veef_refinement}
\begin{tabular}{lccc}
Orbit & $E_{\rm best}^{(8)}$ & $E_{\rm best}^{(16)}$ & $\sigma^{(16)}$ \\
\hline
O09 & 3.321077 & 3.320984 & 0.0809 \\
O12 & 3.068562 & 3.103408 & 0.0398 \\
O13 & 3.349273 & 3.349333 & 0.0346 \\
O15 & 3.401533 & 3.413183 & 0.0423 \\
O19 & 3.484350 & 3.498404 & 0.0330 \\
O20 & 3.504916 & 3.505050 & 0.0476 \\
O21 & 3.198752 & 3.219541 & 0.0842 \\
\end{tabular}
\end{table}

\begin{table*}[tb]
\centering
\caption{Frozen two-time VEEF diagnostic for all 21 reduced rooted orbits. The statistic and threshold were fixed before the full-family extension. Every $L_*=2$ orbit lies below $g_c=0.05$ ebit/$t_0$ and every $L_*=3$ orbit lies above it.}
\label{tab:veef_two_time_full}
\begin{tabular}{lcccc@{\qquad}lcccc}
Orbit & $L_*$ & $E(2t_0)$ & $E(2.4t_0)$ & $g_{2.4}$ & Orbit & $L_*$ & $E(2t_0)$ & $E(2.4t_0)$ & $g_{2.4}$ \\
\hline
O01 & 2 & 3.999792 & 3.999936 & 0.000359 & O12 & 3 & 3.068562 & 3.730865 & 1.655758 \\
O02 & 2 & 3.999576 & 3.999755 & 0.000446 & O13 & 3 & 3.349273 & 3.801060 & 1.129469 \\
O03 & 2 & 3.999629 & 3.999840 & 0.000528 & O14 & 3 & 3.040255 & 3.659606 & 1.548379 \\
O04 & 2 & 3.999595 & 3.999941 & 0.000864 & O15 & 3 & 3.401533 & 3.888087 & 1.216385 \\
O05 & 2 & 3.999904 & 3.999945 & 0.000103 & O16 & 3 & 3.700315 & 3.966747 & 0.666079 \\
O06 & 2 & 3.999530 & 3.999653 & 0.000306 & O17 & 3 & 3.713595 & 3.968033 & 0.636097 \\
O07 & 3 & 3.342932 & 3.815013 & 1.180201 & O18 & 3 & 3.498844 & 3.898209 & 0.998412 \\
O08 & 3 & 3.711661 & 3.997131 & 0.713673 & O19 & 3 & 3.484350 & 3.897752 & 1.033506 \\
O09 & 3 & 3.321077 & 3.836224 & 1.287868 & O20 & 3 & 3.504916 & 3.936694 & 1.079446 \\
O10 & 3 & 3.327647 & 3.803388 & 1.189353 & O21 & 3 & 3.198752 & 3.798042 & 1.498224 \\
O11 & 3 & 3.498990 & 3.899253 & 1.000657 & & & & & \\
\end{tabular}
\end{table*}

\begin{table}[tb]
\centering
\caption{Representative robustness to the VEEF regularization weight. The final column gives the range of $E_{\rm best}$ across the three values of $\lambda$.}
\label{tab:veef_lambda_robustness}
\begin{tabular}{lcccc}
Orbit & $E_{0.025}$ & $E_{0.05}$ & $E_{0.10}$ & range \\
\hline
O06 & 3.999573 & 3.999527 & 3.998794 & 0.000779 \\
O17 & 3.715588 & 3.713590 & 3.681291 & 0.034297 \\
O20 & 3.506815 & 3.504404 & 3.498774 & 0.008041 \\
\end{tabular}
\end{table}

This appendix records the numerical checks underlying Sec.~\ref{sec:veef_fingerprint}. The 256 labeled tree--tree graphs reduce to 21 rooted dynamical orbits under exchange of the two sides. Each orbit is optimized with the same $x$-only VEEF resource, regularization weight $\lambda=0.05$, and eight independent starts at $T=2t_0$. The class label $L_*$ is used only after the VEEF optimization for scoring. Table~\ref{tab:veef_all_orbits} lists the single-time results.

The pre-specified single-time rule $E_{\rm best}>3.70\Rightarrow L_*=2$ gives $18/21$ correct classifications, with O08, O16, and O17 as the only exceptions. The class score sets themselves do not overlap. Quantitatively,
\begin{align}
\Delta E_{\rm full}^{\rm best}
&\equiv\min_{L_*=2}E_{\rm VEEF}(2t_0;G)
-\max_{L_*=3}E_{\rm VEEF}(2t_0;G)\nonumber\\
&=0.28594\ \text{ebit}.
\label{eq:veef_full_gap}
\end{align}
The largest eight-start standard deviation is $\sigma_{\max}=0.06332$ ebit, so this post-scoring gap does not meet the conservative pre-specified $5\sigma_{\max}$ criterion. For context, the multi-start mean lies above $3.70$ for all six $L_*=2$ orbits and below $3.70$ for all 15 $L_*=3$ orbits. This post-scoring observation is not substituted for the pre-specified best-start rule.

To probe numerical stability, we fixed three refinement triggers before the reruns: $|E_{\rm best}-E_{\rm mean}|>0.05$, $\sigma>0.04$, or total runtime above $4000$ s. These criteria select O09, O12, O13, O15, O19, O20, and O21. Repeating those orbits with 16 starts gives Table~\ref{tab:veef_refinement}. Every refined best score remains below $3.70$ and no class assignment flips.

The two-time statistic and threshold were specified before the initial $T_2=2.4t_0$ boundary calculations. That first diagnostic contained all six $L_*=2$ orbits and the three single-time boundary cases. We then froze Eq.~\eqref{eq:veef_two_time_growth}, the threshold $g_c=0.05$ ebit/$t_0$, $\lambda=0.05$, the eight-start protocol, and the random seed before evaluating the remaining 12 $L_*=3$ orbits. Table~\ref{tab:veef_two_time_full} gives the resulting complete 21-orbit test.

The frozen rule identifies all 21 orbits correctly. The $L_*=2$ growth rates lie between $1.03\times10^{-4}$ and $8.64\times10^{-4}$ ebit/$t_0$, while the $L_*=3$ rates lie between $0.636097$ and $1.655758$ ebit/$t_0$. This establishes complete identification of the $L_*=2/3$ partition within the stated 21-orbit symmetry-reduced family.

We also test sensitivity to the VEEF regularization weight on three representative orbits. O06 samples the saturated $L_*=2$ class, O17 samples the single-time boundary family, and O20 samples a slower-replenishing $L_*=3$ orbit. Table~\ref{tab:veef_lambda_robustness} reports the best score at $\lambda=0.025,0.05,0.10$. The score range remains below $0.05$ ebit for each orbit and the ordering O06 $>$ O17 $>$ O20 is preserved for all three values.

This $\lambda$ check is representative rather than orbit exhaustive, so it is used only as a numerical robustness test and not as an independent classification result.


\begin{thebibliography}{99}


\bibitem{Cirac1997}
J.~I. Cirac, P.~Zoller, H.~J. Kimble, and H.~Mabuchi,
``Quantum state transfer and entanglement distribution among distant nodes in a quantum network,''
Phys. Rev. Lett. \textbf{78}, 3221 (1997).

\bibitem{Briegel1998}
H.-J. Briegel, W.~D\"ur, J.~I. Cirac, and P.~Zoller,
``Quantum repeaters: The role of imperfect local operations in quantum communication,''
Phys. Rev. Lett. \textbf{81}, 5932 (1998).

\bibitem{Cirac1999}
J.~I. Cirac, A.~K. Ekert, S.~F. Huelga, and C.~Macchiavello,
``Distributed quantum computation over noisy channels,''
Phys. Rev. A \textbf{59}, 4249 (1999).

\bibitem{Dur2001}
W.~D\"ur, G.~Vidal, J.~I. Cirac, N.~Linden, and S.~Popescu,
``Entanglement capabilities of nonlocal Hamiltonians,''
Phys. Rev. Lett. \textbf{87}, 137901 (2001).

\bibitem{Bennett2003}
C.~H. Bennett, A.~W. Harrow, D.~W. Leung, and J.~A. Smolin,
``On the capacities of bipartite Hamiltonians and unitary gates,''
IEEE Trans. Inf. Theory \textbf{49}, 1895 (2003).

\bibitem{Childs2003}
A.~M. Childs, D.~W. Leung, F.~Verstraete, and G.~Vidal,
``Asymptotic entanglement capacity of the Ising and anisotropic Heisenberg interactions,''
Quantum Inf. Comput. \textbf{3}, 97 (2003).

\bibitem{LiebRobinson1972}
E.~H. Lieb and D.~W. Robinson,
``The finite group velocity of quantum spin systems,''
Commun. Math. Phys. \textbf{28}, 251 (1972).

\bibitem{BravyiHastingsVerstraete2006}
S.~Bravyi, M.~B. Hastings, and F.~Verstraete,
``Lieb--Robinson bounds and the generation of correlations and topological quantum order,''
Phys. Rev. Lett. \textbf{97}, 050401 (2006).

\bibitem{EisertOsborne2006}
J.~Eisert and T.~J. Osborne,
``General entanglement scaling laws from time evolution,''
Phys. Rev. Lett. \textbf{97}, 150404 (2006).

\bibitem{Bravyi2007}
S.~Bravyi,
``Upper bounds on entangling rates of bipartite Hamiltonians,''
Phys. Rev. A \textbf{76}, 052319 (2007).

\bibitem{VanAcoleyen2013}
K.~Van Acoleyen, M.~Mari\"en, and F.~Verstraete,
``Entanglement rates and area laws,''
Phys. Rev. Lett. \textbf{111}, 170501 (2013).

\bibitem{Cubitt2005}
T.~S. Cubitt, F.~Verstraete, and J.~I. Cirac,
``Entanglement flow in multipartite systems,''
Phys. Rev. A \textbf{71}, 052308 (2005).

\bibitem{Khaneja2001}
N.~Khaneja, R.~Brockett, and S.~J. Glaser,
``Time optimal control in spin systems,''
Phys. Rev. A \textbf{63}, 032308 (2001).

\bibitem{Romano2007}
R.~Romano and A.~Del Fabbro,
``Optimal generation of entanglement under local control,''
Phys. Rev. A \textbf{76}, 044302 (2007).

\bibitem{Caneva2009}
T.~Caneva, M.~Murphy, T.~Calarco, R.~Fazio, S.~Montangero,
V.~Giovannetti, and G.~E. Santoro,
``Optimal control at the quantum speed limit,''
Phys. Rev. Lett. \textbf{103}, 240501 (2009).

\bibitem{Koch2022}
C.~P. Koch, U.~Boscain, T.~Calarco, G.~Dirr, S.~Filipp, S.~J. Glaser,
R.~Kosloff, S.~Montangero, T.~Schulte-Herbr\"uggen, D.~Sugny, and F.~K. Wilhelm,
``Quantum optimal control in quantum technologies. Strategic report on current status, visions and goals for research in Europe,''
EPJ Quantum Technol. \textbf{9}, 19 (2022).

\bibitem{Malvetti2024}
E.~Malvetti,
``Entanglement in bipartite quantum systems with fast local unitary control,''
arXiv:2401.07024 (2024).

\bibitem{MalvettiVanDamme2024}
E.~Malvetti and L.~Van Damme,
``Optimal control of bipartite quantum systems,''
arXiv:2405.20034 (2024).

\bibitem{Eldredge2020}
Z.~Eldredge, L.~Zhou, A.~Bapat, J.~R. Garrison, A.~Deshpande,
F.~T. Chong, and A.~V. Gorshkov,
``Entanglement bounds on the performance of quantum computing architectures,''
Phys. Rev. Research \textbf{2}, 033316 (2020).

\bibitem{Bapat2023}
A.~Bapat, A.~M. Childs, A.~V. Gorshkov, and E.~Schoute,
``Advantages and limitations of quantum routing,''
PRX Quantum \textbf{4}, 010313 (2023).

\bibitem{Devulapalli2026}
D.~Devulapalli, C.~Yin, A.~Y. Guo, E.~Schoute, A.~M. Childs, A.~V. Gorshkov, and A.~Lucas,
``Quantum routing and entanglement dynamics through bottlenecks,''
PRX Quantum \textbf{7}, 010310 (2026).

\bibitem{Mishra2015}
S.~K. Mishra, A.~Lakshminarayan, and V.~Subrahmanyam,
``Protocol using kicked Ising dynamics for generating states with maximal multipartite entanglement,''
Phys. Rev. A \textbf{91}, 022318 (2015).

\bibitem{Lu2021}
Y.~Lu, Y.-M. Li, P.-F. Zhou, and S.-J. Ran,
``Preparation of many-body ground states by time evolution with variational microscopic magnetic fields and incomplete interactions,''
Phys. Rev. A \textbf{104}, 052413 (2021).

\bibitem{Lu2024}
Y.~Lu, P.~Shi, X.-H. Wang, J.~Hu, and S.-J. Ran,
``Persistent ballistic entanglement spreading with optimal control in quantum spin chains,''
Phys. Rev. Lett. \textbf{133}, 070402 (2024).

\bibitem{BoteroReznikFermion2004}
A.~Botero and B.~Reznik,
``BCS-like modewise entanglement of fermion Gaussian states,''
Phys. Lett. A \textbf{331}, 39 (2004).

\bibitem{BravyiFermionic2005}
S.~Bravyi,
``Lagrangian representation for fermionic linear optics,''
Quantum Inf. Comput. \textbf{5}, 216 (2005).

\bibitem{Pandey2024}
V.~Pandey, S.~Bhowmick, B.~Mohan, Sohail, and U.~Sen,
``Fundamental speed limits on entanglement dynamics of bipartite quantum systems,''
Phys. Rev. A \textbf{110}, 052420 (2024).

\bibitem{Hamazaki2024}
R.~Hamazaki,
``Speed limits to fluctuation dynamics,''
Commun. Phys. \textbf{7}, 361 (2024).

\bibitem{JordanWigner1928}
P.~Jordan and E.~Wigner,
``{\"U}ber das paulische {\"A}quivalenzverbot,''
Z. Phys. \textbf{47}, 631 (1928).

\bibitem{LiebSchultzMattis1961}
E.~Lieb, T.~Schultz, and D.~Mattis,
``Two soluble models of an antiferromagnetic chain,''
Ann. Phys. (N.Y.) \textbf{16}, 407 (1961).

\bibitem{Hein2004}
M.~Hein, J.~Eisert, and H.~J. Briegel,
``Multiparty entanglement in graph states,''
Phys. Rev. A \textbf{69}, 062311 (2004).

\bibitem{Fattal2004}
D.~Fattal, T.~S. Cubitt, Y.~Yamamoto, S.~Bravyi, and I.~L. Chuang,
``Entanglement in the stabilizer formalism,''
arXiv:quant-ph/0406168 (2004).

\bibitem{Nielsen2003}
M.~A. Nielsen, C.~M. Dawson, J.~L. Dodd, A.~Gilchrist, D.~Mortimer, T.~J. Osborne, M.~J. Bremner, A.~W. Harrow, and A.~Hines,
``Quantum dynamics as a physical resource,''
Phys. Rev. A \textbf{67}, 052301 (2003).

\bibitem{Strang1968}
G.~Strang,
``On the construction and comparison of difference schemes,''
SIAM J. Numer. Anal. \textbf{5}, 506 (1968).

\bibitem{ChildsTrotter2021}
A.~M. Childs, Y.~Su, M.~C. Tran, N.~Wiebe, and S.~Zhu,
``Theory of Trotter error with commutator scaling,''
Phys. Rev. X \textbf{11}, 011020 (2021).

\bibitem{Hirose2018}
M.~Hirose and P.~Cappellaro,
``Time-optimal control with finite bandwidth,''
Quantum Inf. Process. \textbf{17}, 88 (2018).

\bibitem{Fisher2009}
R.~Fisher, H.~Yuan, A.~Sp\"orl, and S.~J. Glaser,
``Time-optimal generation of cluster states,''
Phys. Rev. A \textbf{79}, 042304 (2009).

\bibitem{Langer2026}
M.~Langer, R.~Morral-Yepes, A.~Gammon-Smith, F.~Pollmann, and B.~Kraus,
``Matchgate circuit representation of fermionic Gaussian states: optimal preparation, approximation, and classical simulation,''
arXiv:2603.05675 (2026).

\bibitem{Cabello2011}
A.~Cabello, L.~E. Danielsen, A.~J. L\'opez-Tarrida, and J.~R. Portillo,
``Optimal preparation of graph states,''
Phys. Rev. A \textbf{83}, 042314 (2011).

\bibitem{Kumabe2026}
S.~Kumabe, R.~Mori, and Y.~Yoshimura,
``Complexity of graph-state preparation by Clifford circuits,''
Quantum \textbf{10}, 2165 (2026).

\bibitem{Rao2026}
C.~Rao, H.~Sahu, A.~Bhattacharya, S.~A. Rather, M.~Flory, and Z.~Raissi,
``Graph-State Circuit Blocks control Entanglement and Scrambling Velocities,''
arXiv:2605.11076 (2026).

\end{thebibliography}
\end{document}